\documentclass[lettersize,journal]{IEEEtran}
\usepackage{amsmath,amsfonts}
\usepackage{amssymb}
\usepackage{algorithmic}
\usepackage{algorithm}
\usepackage{array}
\usepackage[caption=false,font=normalsize,labelfont=sf,textfont=sf]{subfig}
\usepackage{textcomp}
\usepackage{stfloats}
\usepackage{url}
\usepackage{verbatim}
\usepackage{pgfplots}
\usepackage{graphicx}
\usepackage{cite}
\usepackage{booktabs}
\usepackage{tikz}
\usetikzlibrary{positioning,backgrounds,fit,arrows.meta}
\usepgfplotslibrary{groupplots}
\pgfplotsset{compat=1.18}

\begin{document}

\title{Recursive Spectral Graph Partitioning Coding for Index Assignment in Semantic Communications}

\author{Leonardo~Roque,~\textit{Student~Member,~IEEE},~Louis-Adrien~Dufrène,~Quentin~Lampin,\\~Grégoire~Lefebvre,~and~Mohamad~Assaad,~\textit{Senior~Member,~IEEE}% <-this % stops a space
\thanks{This research was funded in part by the Agence Nationale de la Recherche (ANR) under the ANR-24-IAS1-0002-02 COMSEMA project and the SNS
JU 6GARROW project under the EU’s Horizon program Grant Agreement
No 101192194, in collaboration with ROK. (\emph{Corresponding author: Leonardo Roque}.)}% <-this % stops a space
\thanks{L. Roque, L.-A. Dufrène, Q. Lampin, and G. Lefebvre are with Orange Research, Meylan, France (e-mail: leonardo.roquealmeidamatos@orange.com; louisadrien.dufrene@orange.com; quentin.lampin@orange.com; gregoire.lefebvre@orange.com).}% <-this % stops a space
\thanks{L. Roque and M. Assaad are with the Laboratoire des Signaux et Systèmes, CentraleSupélec, Gif-sur-Yvette, France (e-mail: mohamad.assaad@centralesupelec.fr).}% <-this % stops a space
}

\maketitle

\begin{abstract}
Transmitting quantized semantic embeddings over a bit-flip channel requires a binary mapping that minimizes expected reconstruction distortion.
The classical Binary Switching Algorithm (BSA) ($\mathcal{O}(K^3 \log K)$ for $K$ centroids) yields locally optimal assignments but ignores hierarchical bit significance, making it incompatible with Unequal Error Protection (UEP).
We propose Recursive Spectral Graph Partitioning (RSGP), which recursively bisects the centroid affinity graph along the Fiedler eigenvector so that early bits separate semantically distant groups while later bits refine local structure.
RSGP runs in $\mathcal{O}(K^3)$ with no iterative outer loop, and is the only method studied to show a systematic positive UEP gain: protecting early bits over later ones improves accuracy for RSGP, while being neutral or harmful for unstructured assignments.
Experiments on the Banking77 dataset and the Massive Text Embedding Benchmark (MTEB) confirm that RSGP consistently matches or exceeds BSA accuracy across codebook sizes and error profiles.
\end{abstract}


\section{Introduction}

Contrary to the traditional communication perspective, semantic communication aims to transmit task-relevant representations of data rather than raw bits, shifting the performance criterion from symbol-level fidelity to downstream task utility~\cite{ansere2026frombt}.
As a structured semantic representation tailored for AI-model integration and heterogeneous use cases, embeddings are high-dimensional vectors that can capture the underlying semantic structure within data.
A common approach to designing semantic systems encodes a data sample as a high-dimensional embedding, compresses it via vector quantization (VQ) into a codebook of $K$ discrete centroids, and transmits the corresponding centroid indices over a digital communication channel~\cite{roque2025pq}.

To make index transmission compatible with digital chains, the choice of a well suited binary mapping---the function $\varphi$ from centroid indices to $m$-bit codewords--- is needed to keep reconstruction quality at reception under channel noise.
This is directly represented and defined by the \emph{Index Assignment Problem} (IAP)~\cite{wu2011iap}:
the expected distortion under independent per-bit transmission errors depends on the Hamming distances between codewords assigned to semantically proximate centroids. A poor assignment causes the most frequent confusion errors to land on semantically distant centroids, degrading both the mean-squared reconstruction error and downstream task accuracy.

As a natural question arising from the IAP study in semantic communication scenarios, one can investigate how the possible binary mapping constructions could intrinsically account for and embed semantic relevance by design. Notably, if task-relevant structure could be directly identified and organized along binary source representations, Unequal Error Protection (UEP) channel coding could be applied to protect, at a higher reliability level, the most significant bits against noise and preserve their conveyed semantic content. A well-known state-of-the-art technique for solving the IAP is the Binary Switching Algorithm (BSA)~\cite{zeger1990pseudogray}, which iteratively swaps pairs of codewords to reduce distortion until convergence to a local minimum.
However, BSA is agnostic to bit-position ordering, producing assignments with no inherent significance hierarchy across bit positions.
Also, when dealing with an increasing number of discrete elements, BSA grows as $\mathcal{O}(K^3 \log K)$ per pass and becomes particularly prohibitive for large $K$ values.

To address both semantic-aware construction and complexity needs, we propose \emph{Recursive Spectral Graph Partitioning} (RSGP), a one-shot index assignment that recursively bisects the centroid affinity graph using the Fiedler eigenvector of the graph Laplacian.
The recursion naturally embeds a coarse-to-fine semantic ordering into the bit positions: the first bit separates the two broadest semantic groups, the second bit refines each group, and so forth.
This hierarchical structure aligns with a UEP protection gradient, and we show both theoretically and empirically that RSGP is the only studied method to achieve a systematic positive UEP gain against the respective uniform error performances.

Our contributions are as follows:
\begin{itemize}
  \item RSGP: a complexity-efficient, semantically hierarchical index assignment growing with $K$ quantization centroids as $\mathcal{O}(K^3)$ with no iterative outer loop.
  \item A complexity comparison of all studied IAP methods.
  \item The generalization and a first evaluation of BSA as a UEP-aware encoding scheme directly addressing IAP.
  \item The analysis of UEP gain as a function of IAP structure: Jensen’s inequality shows that the semantic bit ordering of RSGP ensures that the UEP reduces distortion.
  \item Empirical validation and label classification across different encoding parameters on the Banking77 dataset.
\end{itemize}

\section{System Model}
\label{sec:system_model}

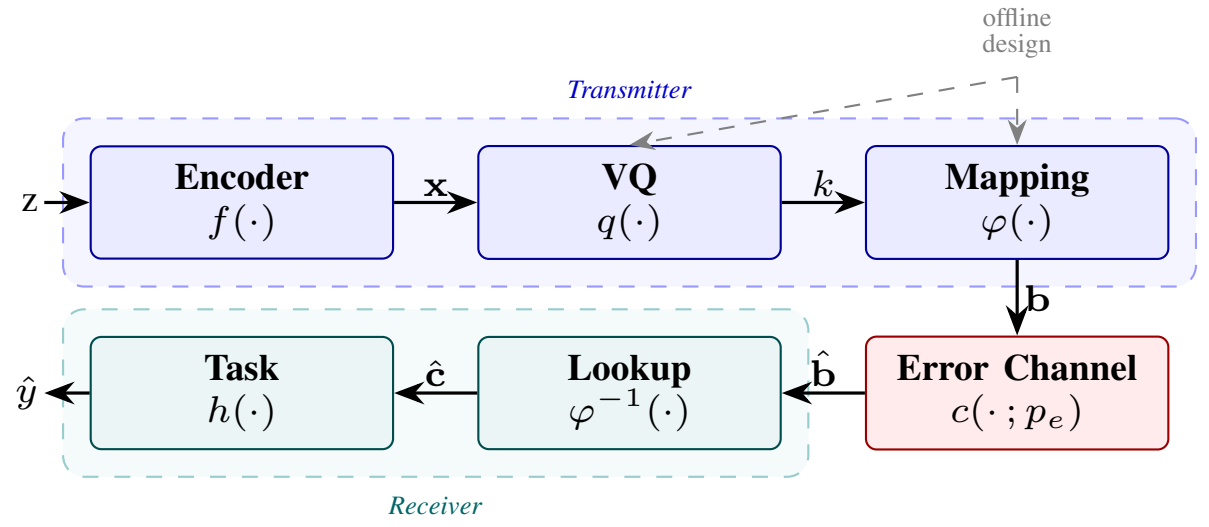
\begin{figure}[!t]
  \centering
  \begin{tikzpicture}[
    tx/.style    = {draw=blue!60!black, fill=blue!8, rounded corners=2pt,
                    minimum width=2cm, minimum height=0.7cm,
                    align=center, font=\scriptsize},
    ch/.style    = {draw=red!60!black, fill=red!8, rounded corners=2pt,
                    minimum width=2cm, minimum height=0.7cm,
                    align=center, font=\scriptsize},
    rx/.style    = {draw=teal!60!black, fill=teal!8, rounded corners=2pt,
                    minimum width=2cm, minimum height=0.7cm,
                    align=center, font=\scriptsize},
    arr/.style   = {->, >=Stealth, semithick},
    siglbl/.style= {font=\scriptsize, inner sep=1pt},
  ]
  %--- Row 1: Transmitter (left to right) ---
  \node[tx] (enc)                     {\textbf{Encoder}\\ $f(\cdot)$};
  \node[tx, right=0.55cm of enc] (vq) {\textbf{VQ}\\ $q(\cdot)$};
  \node[tx, right=0.55cm of vq] (phi) {\textbf{Mapping}\\ $\varphi(\cdot)$};
  %--- Row 2: Channel then Receiver (right to left) ---
  \node[ch, below=0.5cm of phi] (channel)  {\textbf{Error Channel}\\ $c(\cdot\,;p_e)$};
  \node[rx, left=0.55cm of channel] (lkp)  {\textbf{Lookup}\\ $\varphi^{-1}(\cdot)$};
  \node[rx, left=0.55cm of lkp] (task) {\textbf{Task}\\ $h(\cdot)$};
  %--- Input / output nodes ---
  \node[siglbl, left=0.3cm of enc]  (xin)  {z};
  \node[siglbl, left=0.3cm of task] (yout) {$\hat{y}$};
  %--- Arrows with signal labels ---
  \draw[arr] (xin)  -- (enc);
  \draw[arr] (enc)  -- node[siglbl, above]{$\mathbf{x}$} (vq);
  \draw[arr] (vq)   -- node[siglbl, above]{$k$} (phi);
  \draw[arr] (phi)  -- node[siglbl, right]{$\mathbf{b}$} (channel);
  \draw[arr] (channel)  -- node[siglbl, above]{$\hat{\mathbf{b}}$} (lkp);
  \draw[arr] (lkp)  -- node[siglbl, above]{$\hat{\mathbf{c}}$} (task);
  \draw[arr] (task) -- (yout);
  %--- Shaded transmitter / receiver bounding boxes ---
  \begin{scope}[on background layer]
    \node[draw=blue!40, dashed, fill=blue!4, rounded corners=4pt, inner sep=5pt,
          fit=(enc)(vq)(phi),
          label={[font=\tiny,blue!80!black]above:\textit{Transmitter}}] {};
    \node[draw=teal!40, dashed, fill=teal!4, rounded corners=4pt, inner sep=5pt,
          fit=(lkp)(task),
          label={[font=\tiny,teal!80!black]below:\textit{Receiver}}] {};
  \end{scope}
  %--- φ is designed offline, not transmitted ---
  \node[font=\tiny, gray, align=center, above=0.45cm of phi] (ann)
        {offline\\[-1pt]design};
  \draw[->, >=Stealth, gray, thin, dashed] (ann.south) -- (phi.north);
  \draw[->, >=Stealth, gray, thin, dashed] (ann.south) -- (vq.north);
  \end{tikzpicture}
  \caption{Semantic communication system. Both the vector quantizer~$q(\cdot)$ and the binary mapping~$\varphi(\cdot)$ are designed offline; only the $m$-bit codeword~$\mathbf{b}$ is transmitted online over the error channel.}
  \label{fig:system}
\end{figure}

We consider the semantic communication system depicted in Fig.~\ref{fig:system}: a transmitter feeds an embedding model $f(\cdot)$ with a source data sample $z$, encodes the resulting embedding $\mathbf{x}\in\mathbb{R}^d$ into a discrete index $k=q(\mathbf{x})$, and maps it to a codeword $\mathbf{b}=\varphi(k)$ transmitted over a noisy channel.
The receiver reconstructs an approximate vector $\hat{\mathbf{c}}$ from the received codeword $\hat{\mathbf{b}}$, which feeds a task-adapted model $h(\cdot)$ performing the target task, in this work label classification.

\subsection{Quantization, Binary Mapping and Channel Transmission}
The embedding $\mathbf{x}$ is compressed by a vector quantizer ~$q(\cdot)$ into a centroid set $\mathcal{C} = \{\mathbf{c}_1, \ldots, \mathbf{c}_K\} \subset \mathbb{R}^d$ learned in an offline stage, where $K = 2^m$ and $m$ is the number of bits per codeword.
The transmitter maps each embedding to its nearest centroid $    q(\mathbf{x}) = \operatorname*{arg\,min}_{k \in \{1,...,K\}}
                    \|\mathbf{x} - \mathbf{c}_k\|^2,
    \label{eq:quantizer}$
yielding centroid index $k$.
The quantization scheme is obtained by running $K$-Means on a training dataset.

Also learned in an offline stage from the centroid set $\mathcal{C}$, a \emph{binary mapping} is a function $\varphi : \{1,...,K\} \to \{0,1\}^m$ that maps each centroid index to a distinct codeword.
The transmitter sends $\mathbf{b} = \varphi(k)$ over a channel that flips each transmitted bit independently with crossover probability $p_e \in [0,0.5]$. The received codeword is thus given by $\hat{\mathbf{b}} = c(\mathbf{b}; p_e)$, satisfying:
\begin{equation}
    \Pr\!\bigl[\hat{\mathbf{b}} = \varphi(j) \mid \varphi(k) \text{ sent}\bigr]
    = p_e^{H_{kj}} (1 - p_e)^{m - H_{kj}},
    \label{eq:unif-channel}
\end{equation}
where $H_{kj} = d_H\!\bigl(\varphi(k),\varphi(j)\bigr)$ is the Hamming distance.

This uniform, independent bit-flip assumption is motivated by the fact that the residual bit-error pattern delivered to upper layers after physical-layer coding, interleaving, and modulation is typically approximately uniform across the transmitted binary representation in most digital systems, offering a realistic and tractable abstraction that avoids modeling the underlying coding chain explicitly.
Section~\ref{sec:system_model}-C generalizes the uniform error model to a heterogeneous per-bit error profile.

The receiver decodes $\hat{\mathbf{b}}$ by table lookup to obtain the centroid $\mathbf{c}_{\varphi^{-1}(\hat{\mathbf{b}})}$, and the reconstruction distortion induced by channel errors is given by:
\begin{equation}
    D(\varphi) = \sum_{\substack{k,j=1 \\ k \neq j}}^{K}
                 p_e^{H_{kj}}(1-p_e)^{m-H_{kj}}
                 \|\mathbf{c}_k - \mathbf{c}_j\|^2.
    \label{eq:distortion}
\end{equation}
The IAP is to find the binary mapping $\varphi$ that minimizes $D(\varphi)$ for a given centroid set and $p_e$.

\subsection{Downstream Task and Evaluation Metrics}

Performances are characterized along two complementary axes.
At the communication level, the mean squared error (MSE) between the transmitted centroid and its noisy reconstruction quantifies embedding reconstruction fidelity.
To isolate the mapping-induced channel distortion while accounting for the common quantization scheme, we report the normalized metric $\mathrm{MSE}_{\mathrm{norm}}(p_e) =
\frac{\mathrm{MSE}(p_e)}{\mathrm{MSE}_{\mathrm{quant}}}$, where $\mathrm{MSE}(p_e) = \frac{1}{N_{\mathrm{eval}}}
  \sum_{j=1}^{N_{\mathrm{eval}}}
  \bigl\|\mathbf{c}_{q(\mathbf{x}_j)} - \hat{\mathbf{c}}_j\bigr\|^2,$  and $\mathrm{MSE}_{\mathrm{quant}} =
  \frac{1}{N_{\mathrm{eval}}}\sum_{j=1}^{N_{\mathrm{eval}}}
  \min_k\|\mathbf{x}_j-\mathbf{c}_k\|^2$.
Here, $\mathrm{MSE}_{\mathrm{quant}}$ is common to all mappings for a fixed codebook and provides the quantization-only reference scale.

At the task level, we measure the semantic relevance of the received embeddings through an intent-classification pipeline.
Under the Massive Text Embedding Benchmark (MTEB) protocol~\cite{muennighoff2023mteb}, a logistic regression classifier $h : \mathbb{R}^d \to \mathcal{Y}$ is trained on uncompressed training embeddings, then applied to the received centroids $\{\hat{\mathbf{c}}_j\}$ during online transmission.
The service-centric metric is the classification accuracy, defined as the ratio between correct and total number of predictions. Similarly to MSE, we isolate the mapping-induced observed performance from the fixed quantization by reporting the normalized accuracy $\mathrm{Acc}_{\mathrm{norm}}(p_e) = \frac{\mathrm{Acc}(p_e)}{\mathrm{Acc}(p_e=0)}$, where $\mathrm{Acc}(p_e=0)$ is the accuracy obtained with the same quantizer and mapping when the channel is error-free.
MSE is retained as an embedding reconstruction capacity metric, while $\mathrm{Acc}$ constitutes the primary task-adapted measure for the studied semantic communication system.

\subsection{Unequal Error Protection}

In a UEP scheme, each bit position $\ell \in \{0,\ldots,m-1\}$ has an independent crossover probability $p_\ell$, giving a per-bit profile $\mathbf{p} = (p_0,\ldots,p_{m-1})$.
The transition probability in~\eqref{eq:unif-channel} generalizes to:
\begin{equation}
    \Pr\!\bigl[\hat{\mathbf{b}}=\varphi(j)\mid\varphi(k)\bigr]
    = \prod_{\ell=0}^{m-1}
      p_\ell^{b_\ell^{(k)}\oplus b_\ell^{(j)}}
      (1-p_\ell)^{1-b_\ell^{(k)}\oplus b_\ell^{(j)}},
    \label{eq:uep_transition}
\end{equation}
where $b_\ell^{(k)}$ is the $\ell$-th bit of $\varphi(k)$, and $\oplus$ represents the XOR logical operator.
We measure the \emph{normalized UEP gain} as the improvement in normalized downstream accuracy when the profile $\mathbf{p}$ is applied versus the uniform error model applied across all bit positions at a fixed mean $\bar{p} = \frac{1}{m}\sum_\ell p_\ell$:
\begin{equation}
    \Delta_{\mathrm{UEP}}(\varphi)
    = \mathrm{Acc}_{\mathrm{norm}}(\varphi, \mathbf{p})
    - \mathrm{Acc}_{\mathrm{norm}}(\varphi, \bar{p}\,\mathbf{1}_m),
    \label{eq:uep_gain}
\end{equation}
where $\mathbf{1}_m$ represents the all-ones vector of dimension $m$.

\section{Index Assignment Problem and Baselines}
\label{sec:baselines}

Minimizing $D(\varphi)$ exactly requires searching over all $K!$ binary mappings---a space that already exceeds $2\times10^{13}$ possible mappings for $K=16$ ($m=4$).
The IAP belongs to the family of quadratic assignment problems and is NP-hard in general~\cite{wu2011iap}, so practical algorithms resort to local search heuristics that converge to a local minimum of $D(\varphi)$ from a given starting point.
We compare RSGP against two such references.

\textbf{Random} assigns the $K$ binary codewords to centroids via a uniformly random permutation drawn independently of the semantic embedding geometry within the quantization centroids.
By symmetry, every random binary mapping achieves the same expected distortion $\mathbb{E}[D(\varphi_\mathrm{rand})]$, serving as a lower reference bound for the case in which no semantic structure is considered for index assignment.

\textbf{BSA}~\cite{zeger1990pseudogray} starts from a random initial assignment and repeatedly scans all $\binom{K}{2}$ candidate pairwise code swaps, applying each swap immediately whenever it strictly reduces $D(\varphi)$. Since $D(\varphi)$ strictly decreases at each accepted swap over a finite set of mappings, BSA is guaranteed to converge to a local minimum whose quality depends on the random initialization.
Notably, BSA optimizes $D(\varphi)$ for a single fixed $p_e$ and treats all $m$ bit positions equivalently, imposing no semantic-aware ordering reflecting the embedding geometry.

\textbf{UEP-BSA} is introduced as a direct generalization of the BSA baseline to the UEP error profile $\mathbf{p}$. The uniform transition probability in~\eqref{eq:unif-channel} is replaced by the expression in~\eqref{eq:uep_transition}. UEP-BSA then applies the same sequential pairwise-swap rule to minimize the resulting profile-aware distortion.

\section{Recursive Spectral Graph Partitioning}
\label{sec:method}
We construct a weighted, fully connected, undirected affinity graph $\mathcal{G} = (\mathcal{C}, S)$ whose nodes are the centroids of the quantization set $\mathcal{C}$ and whose edge weights $S_{ij}$, defined in~\eqref{eq:affinity}, increase with the pairwise proximity of $\mathbf{c}_i$ and $\mathbf{c}_j$ in the embedding space. We use this geometric proximity as a proxy for task-relevant semantic similarity between centroids, motivated by the fact that downstream task accuracy degrades with reconstruction distortion in the embedding space. Index assignment can then be cast as a \emph{recursive graph partitioning problem}: we seek $m$ nested bipartitions of $\mathcal{G}$ such that strongly connected nodes---centroid pairs with high affinity---are more likely to remain grouped together throughout the recursion, and therefore receive codewords at small Hamming distance.

The binary structure of a codeword admits a natural interpretation within this recursive process: at each bit position $\ell$, every centroid within a given subgroup is placed on one side of a bipartition of that subgroup, and bit at position $\ell$ of its codeword records which side it was assigned to.
The critical observation is that the depth at which two centroids are first separated determines the significance of their distinguishing bit. Thus, the bit index acts as a proxy for partition granularity, with coarse structure encoded in the first and most significant bits and fine-grained local structure in the last and least significant ones. This coarse-to-fine ordering is naturally exploited by a UEP profile configured with monotonically decreasing protection along bit depth: by construction, RSGP aligns such a profile with its affinity-based bipartition hierarchy, whereas unstructured assignments carry no specific ordering.

\subsection{Algorithm}

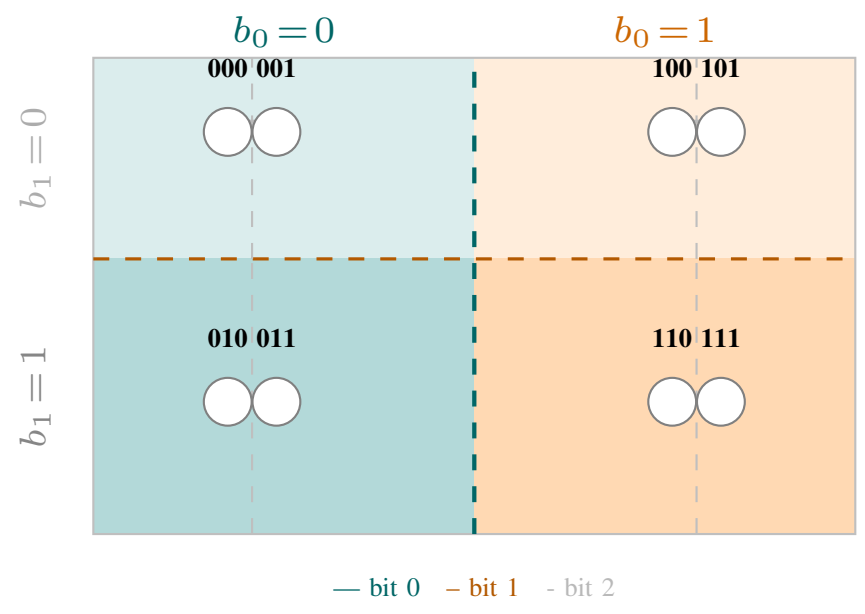
\begin{figure}[!t]
  \centering
  \begin{tikzpicture}[scale=0.7,
    cpt/.style={circle, draw=black!50, fill=white, minimum size=9pt, inner sep=0pt},
  ]
    %--- Quadrant backgrounds: teal for bit0=0, orange for bit0=1 ---
    \fill[teal!14]   (0.0,2.6) rectangle (3.6,4.5);
    \fill[teal!30]   (0.0,0.0) rectangle (3.6,2.6);
    \fill[orange!14] (3.6,2.6) rectangle (7.2,4.5);
    \fill[orange!30] (3.6,0.0) rectangle (7.2,2.6);
    \draw[black!25, thin] (0,0) rectangle (7.2,4.5);
    %--- Bisection lines: level 0 (teal thick), 1 (orange medium), 2 (gray thin) ---
    \draw[teal!80!black, thick, dashed] (3.6,0) -- (3.6,4.5);
    \draw[orange!70!black, semithick, dashed] (0,2.6) -- (7.2,2.6);
    \draw[gray!50, thin, dashed] (1.5,2.6) -- (1.5,4.5);
    \draw[gray!50, thin, dashed] (1.5,0.0) -- (1.5,2.6);
    \draw[gray!50, thin, dashed] (5.7,2.6) -- (5.7,4.5);
    \draw[gray!50, thin, dashed] (5.7,0.0) -- (5.7,2.6);
    %--- Affinity edges: nearest-neighbour (bit-2) pairs ---
    \draw[gray!55, semithick] (1.3,3.9) -- (1.7,3.9);
    \draw[gray!55, semithick] (1.3,1.25) -- (1.7,1.25);
    \draw[gray!55, semithick] (5.5,3.9) -- (5.9,3.9);
    \draw[gray!55, semithick] (5.5,1.25) -- (5.9,1.25);
    %--- Centroids with codeword labels ---
    \node[cpt] (n000) at (1.27,3.8) {};  \node[font=\tiny\bfseries, above=2pt of n000] {000};
    \node[cpt] (n001) at (1.73,3.8) {};  \node[font=\tiny\bfseries, above=2pt of n001] {001};
    \node[cpt] (n010) at (1.27,1.25) {}; \node[font=\tiny\bfseries, above=2pt of n010] {010};
    \node[cpt] (n011) at (1.73,1.25) {}; \node[font=\tiny\bfseries, above=2pt of n011] {011};
    \node[cpt] (n100) at (5.47,3.8) {};  \node[font=\tiny\bfseries, above=2pt of n100] {100};
    \node[cpt] (n101) at (5.93,3.8) {};  \node[font=\tiny\bfseries, above=2pt of n101] {101};
    \node[cpt] (n110) at (5.47,1.25) {}; \node[font=\tiny\bfseries, above=2pt of n110] {110};
    \node[cpt] (n111) at (5.93,1.25) {}; \node[font=\tiny\bfseries, above=2pt of n111] {111};
    %--- Region labels ---
    \node[teal!80!black, font=\scriptsize] at (1.8,4.75) {$b_0\!=\!0$};
    \node[orange!80!black, font=\scriptsize] at (5.4,4.75) {$b_0\!=\!1$};
    \node[gray!65, font=\scriptsize, rotate=90] at (-0.55,3.55) {$b_1\!=\!0$};
    \node[gray!95, font=\scriptsize, rotate=90] at (-0.55,1.30) {$b_1\!=\!1$};
    %--- Legend ---
    \node[align=center, font=\tiny] at (3.6,-0.5)
      {\textcolor{teal!80!black}{--- bit 0}\quad
       \textcolor{orange!70!black}{-- bit 1}\quad
       \textcolor{gray!55}{- bit 2}};
  \end{tikzpicture}
  \caption{Illustrative example of RSGP on $K=8$ centroids ($m=3$ bits). Dashed lines show the three successive Fiedler cuts.}
  \label{fig:rsgp_illustration}
\end{figure}

Given the centroid set $\mathcal{C}$ and bandwidth $\sigma$ (in this work, set to the median pairwise distance between centroids in $\mathcal{C}$ by default), we build the Gaussian affinity matrix:
\begin{equation}
    S_{ij} = \exp\!\Bigl(-\tfrac{\|\mathbf{c}_i - \mathbf{c}_j\|^2}{2\sigma^2}\Bigr),
    \quad i \neq j, \quad S_{ii}=0.
    \label{eq:affinity}
\end{equation}
From $S$ structuring the relationships between graph nodes, the unnormalized graph Laplacian is given by $L = G - S$, where $G$ is the diagonal degree matrix whose $i$-th entry is $\sum_j S_{ij}$.
The \emph{Fiedler eigenvector}~\cite{fiedler1973}---the eigenvector associated with the second smallest eigenvalue $\lambda_2(L)$ of the graph Laplacian matrix---provides the optimal solution for the continuous relaxation of the minimum graph cut problem~\cite{von2007tutorial}, preserving high affinity nodes in the same partitioned subset.
For each bipartition level, we sort centroids by their Fiedler projected coordinate and split the sorted list at its median value, assigning bit~$0$ to the lower half and bit~$1$ to the upper half.
The procedure recurses independently on each half for the next bit position, terminating after $m = \log_2 K$ levels.
Particularly, for $K$ a power of $2$, this recursive coding structure creates equal partition sizes at every level, which guaranties a valid bijection throughout the recursion.
Algorithm~\ref{alg:rsgp} resumes the RSGP coding implementation, while Fig.~\ref{fig:rsgp_illustration} illustrates its construction on $K=8$ centroids ($m=3$).

\begin{algorithm}[!t]
\caption{RSGP: Recursive Spectral Graph Partitioning}
\label{alg:rsgp}
\footnotesize
\begin{algorithmic}
\REQUIRE Centroids $\mathcal{C}$, bit depth $m$, bandwidth $\sigma$
\STATE Compute $S$ via~\eqref{eq:affinity}; set \texttt{indices} $\leftarrow \{1,...,K\}$, \texttt{bit} $\leftarrow 0$
\STATE \textbf{function} \textsc{Bisect}(\texttt{indices}, \texttt{bit}):
\STATE \quad \textbf{if} \texttt{bit} $= m$ \textbf{then return} \hfill\COMMENT{all $m$ bits written}
\STATE \quad Compute $G$, then $L$ on sub-graph induced by \texttt{indices}
\STATE \quad Compute $\mathbf{f}$ as the Fiedler eigenvector of $L$
\STATE \quad Sort \texttt{indices} by $\mathbf{f}$; split into halves $A$, $B$ at the median value of $\mathbf{f}$
\STATE \quad Set $\varphi_{\texttt{bit}}(i) \leftarrow 0$ for $i\in A$; $\varphi_{\texttt{bit}}(i) \leftarrow 1$ for $i\in B$
\STATE \quad \textsc{Bisect}($A$, \texttt{bit}$+1$); \textsc{Bisect}($B$, \texttt{bit}$+1$)
\STATE \textbf{end function};
\end{algorithmic}
\end{algorithm}

\subsection{Analysis on UEP Adaptation}
\label{sec:uep-alignment}
The recursive structure of RSGP creates a natural \emph{bit-significance hierarchy}: bit position~$0$ is set by the root bisection, which splits the centroids into the two most distinct groups in terms of the affinity weights reflecting embedding proximity. Therefore, pairs differing at bit~$0$ have, on average, the largest pairwise squared distance $\|\mathbf{c}_k - \mathbf{c}_j\|^2$; each subsequent level refines within smaller and more homogeneous subgroups, so bit~$m-1$ separates only the finest nearest-neighbor pairs.

First, let us assume an UEP profile $\mathbf{p}$ configured with a monotonically decreasing protection gradient defined by $p_0 < p_1 < \cdots < p_{m-1}$ .
For a pair $(k,j)$ with $H_{kj}=h$ differing bit positions $I_{kj} = \{\ell : b_\ell^{(k)} \oplus b_\ell^{(j)} = 1\}$, the confusion probability in generating error-induced codewords under the UEP profile~$\mathbf{p}$ is $W_{kj}(\mathbf{p})
  = \prod_{\ell\in I_{kj}}\!p_\ell
    \cdot\!\prod_{\ell\notin I_{kj}}\!(1-p_\ell).
  \label{eq:Wkj}$
Taking logarithms decomposes $\log W_{kj}$ into two sums:
\begin{equation}
  \log W_{kj}(\mathbf{p})
  = \underbrace{\sum_{\ell\in I_{kj}}\!\log p_\ell}_{\text{differing bits}}
  + \underbrace{\sum_{\ell\notin I_{kj}}\!\log(1-p_\ell)}_{\text{matching bits}}.
  \label{eq:logWkj}
\end{equation}
Since $\log(\cdot)$ is strictly concave, Jensen's inequality 
gives:
\begin{align}
  \frac{1}{h}\sum_{\ell\in I_{kj}}\!\log p_\ell
    &\leq \log\!\left(\frac{1}{h}\sum_{\ell\in I_{kj}} p_\ell\right)
     = \log \bar{p}_{I_{kj}}, \label{eq:jensen1}\\
  \frac{1}{m-h}\sum_{\ell\notin I_{kj}}\!\log(1-p_\ell)
    &\leq \log\!\left(1 - \frac{1}{m-h}\sum_{\ell\notin I_{kj}} p_\ell\right) \\
    &= \log(1-\bar{p}_{I^c_{kj}}), \label{eq:jensen2}
\end{align}
where $\bar{p}_{I_{kj}}$ and $\bar{p}_{I^c_{kj}}$ are the mean error rates over
the differing and matching bit positions, respectively.
Substituting into~\eqref{eq:logWkj} and exponentiating, we find: 
\begin{equation}
  W_{kj}(\mathbf{p})
  \;\leq\;
  \bar{p}_{I_{kj}}^{\,h}\;(1-\bar{p}_{I^c_{kj}})^{m-h}.
  \label{eq:jensen_log}
\end{equation}

For RSGP, the most semantically distant pairs (large $\|\mathbf{c}_k-\mathbf{c}_j\|^2$,
highest distortion contribution) have $I_{kj}$ concentrated at the smallest-$\ell$
positions---the most protected bits via the considered UEP profile $\mathbf{p}$.
Consequently, $\bar{p}_{I_{kj}} \leq \bar{p}$ and $\bar{p}_{I^c_{kj}} \geq \bar{p}$,
so both factors in~\eqref{eq:jensen_log} satisfy
$\bar{p}_{I_{kj}} \leq \bar{p}$ and $1-\bar{p}_{I^c_{kj}} \leq 1-\bar{p}$.
Therefore, $W_{kj}^{\text{UEP}}
  \leq \bar{p}_{I_{kj}}^{\,h}\,(1-\bar{p}_{I^c_{kj}})^{m-h}
  \leq \bar{p}^{\,h}\,(1-\bar{p})^{m-h}
  = W_{kj}^{\text{unif}}.
  \label{eq:wkj_rsgp}$
For the most semantically damaging pairs, RSGP's UEP profile
strictly suppresses the confusion probability below the
uniform error baseline.

For unstructured assignments such as Random and BSA,
$I_{kj}$ is independent of $\|\mathbf{c}_k-\mathbf{c}_j\|^2$:
distant and nearby pairs alike have
$\bar{p}_{I_{kj}} \approx \bar{p}$, so the bound~\eqref{eq:jensen_log}
gives $W_{kj}^{\text{UEP}} \approx W_{kj}^{\text{unif}}$ for all pairs.
In this case, UEP merely redistributes the confusion probability without
reducing it where it causes the most harmful damage.

\section{Experimental Results}
\label{sec:experiments}

We evaluate on the Banking77 intent-detection dataset\footnote{\url{https://huggingface.co/datasets/mteb/banking77}}, which consists of customer service queries from the banking domain, each labeled with one of $77$ different user intents (e.g., \emph{card arrival}, \emph{transfer not received by the recipient}, \emph{exchange rate}).
Each query is encoded using $768$-dimensional sentence embeddings from
the fixed pre-trained model
\texttt{tomaarsen/mpnet-base-nli}\footnote{\url{https://huggingface.co/tomaarsen/mpnet-base-nli}}.

Classification accuracy follows the MTEB protocol~\cite{muennighoff2023mteb}: a logistic regression classifier is trained on at most 8 samples per class drawn from the training split, then evaluated on reconstructed test embeddings.
The results are averaged over ten independent MTEB experiments, with different training subsampling and error channel realization used during evaluation.
For the UEP evaluation, we use the fixed-ratio profile $p_\ell = \cfrac{p_{\max}}{\mathrm{ratio}^{m-1-\ell}}$ with $\mathrm{ratio}=10$ to simulate a progressive and monotonically decreasing protection gradient defined in Sec.~\ref{sec:uep-alignment}, sweeping $p_{\max}$ over 8 log-spaced values.
The x-axis of UEP plots shows the mean $\bar{p}$ for a fair comparison with the uniform error baseline.

For each $m \in \{8,10,12\}$, we fit K-Means for $K=2^m$ representative centroids with $n_{\mathrm{init}}=5$ different initializations and build the binary mappings for the same set of obtained centroids using each studied method.
For both Random and BSA, we generate 10 independent binary initializations. Random retains the assignment with the lowest computed distortion at $p_e=0.1$, while BSA starts from each random permutation and performs the sequential distortion evaluation procedure until convergence, with at most 200 full scans; the BSA mapping with the lowest final distortion is retained. We additionally evaluate UEP-BSA, which applies the same sequential procedure with a fixed-ratio profile $p_\ell$ at $p_{\max}=0.1$. RSGP follows the presented Algorithm~\ref{alg:rsgp}.

\subsection{Performance Evaluation}

\pgfplotsset{
  % --- RSGP (teal) ---
  rsgp10/.style={teal!80!black, thick, dashed,
                 mark=*, mark size=1.3pt, mark options={fill=teal!80!black}, mark repeat=2},
  rsgp12/.style={teal!80!black, thick, densely dotted,
                 mark=square*, mark size=1.2pt, mark options={fill=teal!80!black}, mark repeat=2},
  rsgp13/.style={teal!80!black, thick, dash dot,
                 mark=triangle*, mark size=1.6pt, mark options={fill=teal!80!black}, mark repeat=2},
  % --- BSA (orange) ---
  bsa10/.style={orange!80!black, thick, dashed,
                mark=*, mark size=1.3pt, mark options={fill=orange!80!black}, mark repeat=2},
  bsa12/.style={orange!80!black, thick, densely dotted,
                mark=square*, mark size=1.2pt, mark options={fill=orange!80!black}, mark repeat=2},
  bsa13/.style={orange!80!black, thick, dash dot,
                mark=triangle*, mark size=1.6pt, mark options={fill=orange!80!black}, mark repeat=2},
  % --- Random (gray) ---
  rand10/.style={black!45, thin, dashed,
                 mark=*, mark size=1.1pt, mark options={fill=black!45}, mark repeat=2},
  rand12/.style={black!45, thin, densely dotted,
                 mark=square*, mark size=1.0pt, mark options={fill=black!45}, mark repeat=2},
  rand13/.style={black!45, thin, dash dot,
                 mark=triangle*, mark size=1.3pt, mark options={fill=black!45}, mark repeat=2},
  % --- UEP-BSA (purple) ---
  uep10/.style={purple!70!black, thin, dashed,
                mark=*, mark size=1.1pt, mark options={fill=purple!70!black}, mark repeat=2},
  uep12/.style={purple!70!black, thin, densely dotted,
                mark=square*, mark size=1.0pt, mark options={fill=purple!70!black}, mark repeat=2},
  uep13/.style={purple!70!black, thin, dash dot,
                mark=triangle*, mark size=1.3pt, mark options={fill=purple!70!black}, mark repeat=2},
}

\begin{figure}[!t]
  \centering
  \begin{tikzpicture}
  \begin{groupplot}[
    group style={
      group size=1 by 2,
      group name=fig,
      vertical sep=1.2cm,
    },
    width=\columnwidth, height=4.38cm,
    xmode=log,
    grid=both,
    grid style={line width=0.3pt, draw=gray!25},
    major grid style={line width=0.5pt, draw=gray!40},
    tick label style={font=\scriptsize},
    label style={font=\scriptsize},
  ]

  %=================================================================
  % SUBPLOT (a): Accuracy vs p_e
  %=================================================================
  \nextgroupplot[
    xlabel={\shortstack{Crossover probability $p_e$\\[0.00001pt]\textbf{(a)}}},
    xlabel style={align=center},
    ylabel={Normalized accuracy (\%)},
    yticklabel style={text width=1.8em, align=right},
    xmin=5e-5, xmax=0.5, ymin=0, ymax=105,
    legend to name=sharedlegend,
    legend columns=3,
    legend style={font=\tiny, draw=gray!40,
                  /tikz/every even column/.append style={column sep=0.25cm}},
    legend cell align=left,
  ]
  \addplot[rsgp10, mark repeat=1] coordinates{(0.00010,99.92)(0.00022,99.82)(0.00048,99.50)(0.00105,99.06)
    (0.00229,97.88)(0.00500,95.65)(0.01093,90.30)(0.02391,79.20)
    (0.05228,60.26)(0.11433,32.07)(0.25000,8.16)(0.29163,5.41)
    (0.34020,3.38)(0.39685,2.49)(0.46294,1.86)};
  \addlegendentry{RSGP, $m=10$}
  \addplot[rsgp12, mark repeat=1] coordinates{(0.00010,99.91)(0.00022,99.84)(0.00048,99.47)(0.00105,98.81)
    (0.00229,97.53)(0.00500,94.49)(0.01093,88.46)(0.02391,75.96)
    (0.05228,55.24)(0.11433,27.26)(0.25000,6.32)(0.29163,4.25)
    (0.34020,2.75)(0.39685,2.05)(0.46294,1.70)};
  \addlegendentry{RSGP, $m=12$}
  \addplot[rsgp13, mark repeat=1] coordinates{(0.00010,99.84)(0.00022,99.78)(0.00048,99.50)(0.00105,98.70)
    (0.00229,97.48)(0.00500,94.42)(0.01093,87.95)(0.02391,74.84)
    (0.05228,53.98)(0.11433,26.48)(0.25000,5.56)(0.29163,3.88)
    (0.34020,2.64)(0.39685,1.99)(0.46294,1.62)};
  \addlegendentry{RSGP, $m=13$}
  \addplot[bsa10, mark repeat=1] coordinates{(0.00010,99.95)(0.00022,99.74)(0.00048,99.55)(0.00105,98.97)
    (0.00229,97.74)(0.00500,95.06)(0.01093,89.91)(0.02391,78.48)
    (0.05228,59.57)(0.11433,30.90)(0.25000,7.13)(0.29163,4.45)
    (0.34020,2.96)(0.39685,2.28)(0.46294,1.80)};
  \addlegendentry{BSA, $m=10$}
  \addplot[bsa12, mark repeat=1] coordinates{(0.00010,99.91)(0.00022,99.76)(0.00048,99.38)(0.00105,98.73)
    (0.00229,97.27)(0.00500,94.17)(0.01093,87.85)(0.02391,74.93)
    (0.05228,53.35)(0.11433,24.55)(0.25000,4.75)(0.29163,3.13)
    (0.34020,2.26)(0.39685,1.76)(0.46294,1.71)};
  \addlegendentry{BSA, $m=12$}
  \addplot[bsa13, mark repeat=1] coordinates{(0.00010,99.85)(0.00022,99.72)(0.00048,99.33)(0.00105,98.61)
    (0.00229,97.02)(0.00500,93.90)(0.01093,86.76)(0.02391,73.63)
    (0.05228,50.36)(0.11433,22.14)(0.25000,3.99)(0.29163,2.74)
    (0.34020,2.15)(0.39685,1.76)(0.46294,1.74)};
  \addlegendentry{BSA, $m=13$}
  \addplot[rand10, mark repeat=1] coordinates{(0.00010,99.86)(0.00022,99.76)(0.00048,99.41)(0.00105,98.91)
    (0.00229,97.62)(0.00500,95.34)(0.01093,89.30)(0.02391,79.06)
    (0.05228,59.51)(0.11433,30.58)(0.25000,7.26)(0.29163,4.62)
    (0.34020,3.22)(0.39685,2.08)(0.46294,1.60)};
  \addlegendentry{Rand., $m=10$}
  \addplot[rand12, mark repeat=1] coordinates{(0.00010,99.86)(0.00022,99.75)(0.00048,99.42)(0.00105,98.79)
    (0.00229,97.47)(0.00500,94.21)(0.01093,87.79)(0.02391,75.25)
    (0.05228,53.46)(0.11433,24.44)(0.25000,4.79)(0.29163,3.22)
    (0.34020,2.11)(0.39685,1.84)(0.46294,1.68)};
  \addlegendentry{Rand., $m=12$}
  \addplot[rand13, mark repeat=1] coordinates{(0.00010,99.87)(0.00022,99.75)(0.00048,99.43)(0.00105,98.65)
    (0.00229,97.02)(0.00500,93.87)(0.01093,86.32)(0.02391,73.05)
    (0.05228,50.37)(0.11433,21.87)(0.25000,4.06)(0.29163,2.74)
    (0.34020,2.04)(0.39685,1.87)(0.46294,1.64)};
  \addlegendentry{Rand., $m=13$}
  \addplot[uep10, mark repeat=1] coordinates{(0.00010,99.88)(0.00022,99.80)(0.00048,99.55)(0.00105,98.95)
    (0.00229,97.83)(0.00500,95.23)(0.01093,89.79)(0.02391,78.59)
    (0.05228,59.21)(0.11433,30.97)(0.25000,7.14)(0.29163,4.93)
    (0.34020,2.88)(0.39685,2.06)(0.46294,1.77)};
  \addlegendentry{UEP-BSA, $m=10$}
  \addplot[uep12, mark repeat=1] coordinates{(0.00010,99.92)(0.00022,99.67)(0.00048,99.41)(0.00105,98.71)
    (0.00229,97.37)(0.00500,94.38)(0.01093,87.72)(0.02391,74.91)
    (0.05228,53.12)(0.11433,24.02)(0.25000,4.64)(0.29163,2.87)
    (0.34020,2.28)(0.39685,1.89)(0.46294,1.68)};
  \addlegendentry{UEP-BSA, $m=12$}
  \addplot[uep13, mark repeat=1] coordinates{(0.00010,99.88)(0.00022,99.72)(0.00048,99.45)(0.00105,98.67)
    (0.00229,97.11)(0.00500,93.59)(0.01093,87.52)(0.02391,73.64)
    (0.05228,50.00)(0.11433,22.03)(0.25000,3.75)(0.29163,2.72)
    (0.34020,2.00)(0.39685,1.62)(0.46294,1.59)};
  \addlegendentry{UEP-BSA, $m=13$}
  \label{fig:accuracy}

  % --- Zoom-box corners in mainAxis data coordinates ---
  \coordinate (zoomTL) at (axis cs:0.008,92);
  \coordinate (zoomBR) at (axis cs:0.06,45);
  
  %=================================================================
  % SUBPLOT (b): UEP gain vs mean BER
  %=================================================================
  \nextgroupplot[
    xlabel={\shortstack{Mean BER $\bar{p}$\\[0.00001pt]\textbf{(b)}}},
    xlabel style={align=center},
    ylabel={$\Delta_{\mathrm{UEP}}$ (normalized pp)},
    yticklabel style={text width=1.8em, align=right},
    xmin=8e-5, xmax=0.07, ymin=-10, ymax=9.5,
  ]
  \addplot[black!35, thin, forget plot] coordinates{(8e-5,0)(0.07,0)};
  \addplot[rsgp10, mark repeat=1] coordinates{(0.000111,0.008)(0.000266,-0.016)(0.000637,0.115)(0.001524,0.214)
    (0.003647,0.804)(0.008728,1.163)(0.020890,1.596)(0.050000,1.229)};
  \addplot[rsgp12, mark repeat=1] coordinates{(0.0000926,0.000)(0.000222,0.061)(0.000530,0.090)(0.001270,0.258)
    (0.003039,0.598)(0.007273,1.617)(0.017408,3.222)(0.041667,3.123)};
  \addplot[rsgp13, mark repeat=1] coordinates{(0.0000855,0.024)(0.000205,0.069)(0.000490,0.139)(0.001172,0.575)
    (0.002805,1.101)(0.006714,2.300)(0.016069,4.535)(0.038462,8.377)};
  \addplot[bsa10, mark repeat=1] coordinates{(0.000111,-0.099)(0.000266,0.008)(0.000637,0.070)(0.001524,-0.016)
    (0.003647,0.259)(0.008728,-0.449)(0.020890,-1.112)(0.050000,-7.122)};
  \addplot[bsa12, mark repeat=1] coordinates{(0.0000926,0.086)(0.000222,-0.058)(0.000530,-0.078)(0.001270,0.037)
    (0.003039,0.181)(0.007273,-0.473)(0.017408,-1.859)(0.041667,-6.935)};
  \addplot[bsa13, mark repeat=1] coordinates{(0.0000855,0.020)(0.000205,-0.033)(0.000490,0.025)(0.001172,0.082)
    (0.002805,-0.208)(0.006714,-0.302)(0.016069,-1.287)(0.038462,-7.547)};
  \addplot[rand10, mark repeat=1] coordinates{(0.000111,0.012)(0.000266,0.012)(0.000637,0.062)(0.001524,0.041)
    (0.003647,0.107)(0.008728,-0.252)(0.020890,-0.781)(0.050000,-6.776)};
  \addplot[rand12, mark repeat=1] coordinates{(0.0000926,0.012)(0.000222,-0.057)(0.000530,0.021)(0.001270,-0.156)
    (0.003039,-0.086)(0.007273,-0.090)(0.017408,-1.259)(0.041667,-7.340)};
  \addplot[rand13, mark repeat=1] coordinates{(0.0000855,0.029)(0.000205,-0.045)(0.000490,-0.025)(0.001172,-0.090)
    (0.002805,0.045)(0.006714,-0.503)(0.016069,-1.712)(0.038462,-7.298)};
  \addplot[uep10, mark repeat=1] coordinates{(0.000111,0.058)(0.000266,-0.108)(0.000637,-0.083)(0.001524,-0.137)
    (0.003647,-0.062)(0.008728,-0.207)(0.020890,-1.465)(0.050000,-8.271)};
  \addplot[uep12, mark repeat=1] coordinates{(0.0000926,0.020)(0.000222,-0.020)(0.000530,0.016)(0.001270,0.037)
    (0.003039,0.045)(0.007273,-0.254)(0.017408,-0.982)(0.041667,-7.920)};
  \addplot[uep13, mark repeat=1] coordinates{(0.0000855,0.025)(0.000205,0.054)(0.000490,0.066)(0.001172,0.000)
    (0.002805,0.219)(0.006714,-0.483)(0.016069,-1.574)(0.038462,-7.719)};

  % --- Zoom-box corners for inset (b), covering the three RSGP peaks ---
  \coordinate (zoomTL2) at (axis cs:0.013,9);
  \coordinate (zoomBR2) at (axis cs:0.055,0.7);
  
  \end{groupplot}
  
%=================================================================
  % ZOOM INSET on subplot (a), centered around p_e = 2x10^-2
  %=================================================================
  \begin{axis}[
    name=insetAxis,
    at={(fig c1r1.north east)}, anchor=north east,
    xshift=-3.75cm, yshift=-0.45cm,
    width=2.7cm, height=1.9cm,
    scale only axis=true,
    xmode=log,
    xmin=0.008, xmax=0.06, ymin=45, ymax=92,
    xtick={0.01,0.02,0.05},
    xticklabels={$10^{-2}$,$2{\times}10^{-2}$,$5{\times}10^{-2}$},
    ytick={50,70,90},
    tick label style={font=\tiny},
    axis background/.style={fill=white, fill opacity=0.95},
    axis on top,
    axis line style={draw=gray!70},
  ]
  \addplot[rsgp10, mark repeat=1] coordinates{(0.01093,90.30)(0.02391,79.20)(0.05228,60.26)};
  \addplot[rsgp12, mark repeat=1] coordinates{(0.01093,88.46)(0.02391,75.96)(0.05228,55.24)};
  \addplot[rsgp13, mark repeat=1] coordinates{(0.01093,87.95)(0.02391,74.84)(0.05228,53.98)};
  \addplot[bsa10, mark repeat=1]  coordinates{(0.01093,89.91)(0.02391,78.48)(0.05228,59.57)};
  \addplot[bsa12, mark repeat=1]  coordinates{(0.01093,87.85)(0.02391,74.93)(0.05228,53.35)};
  \addplot[bsa13, mark repeat=1]  coordinates{(0.01093,86.76)(0.02391,73.63)(0.05228,50.36)};
  \addplot[rand10, mark repeat=1] coordinates{(0.01093,89.30)(0.02391,79.06)(0.05228,59.51)};
  \addplot[rand12, mark repeat=1] coordinates{(0.01093,87.79)(0.02391,75.25)(0.05228,53.46)};
  \addplot[rand13, mark repeat=1] coordinates{(0.01093,86.32)(0.02391,73.05)(0.05228,50.37)};
  \addplot[uep10, mark repeat=1]  coordinates{(0.01093,89.79)(0.02391,78.59)(0.05228,59.21)};
  \addplot[uep12, mark repeat=1]  coordinates{(0.01093,87.72)(0.02391,74.91)(0.05228,53.12)};
  \addplot[uep13, mark repeat=1]  coordinates{(0.01093,87.52)(0.02391,73.64)(0.05228,50.00)};
  \end{axis}

  % --- Zoom indicator: dashed box + connector lines ---
  \draw[gray!70, thin, dashed] (zoomTL) rectangle (zoomBR);
  \draw[gray!50, thin] (zoomTL) -- (insetAxis.north west);
  \draw[gray!50, thin] (zoomBR) -- (insetAxis.south east);

  %=================================================================
  % ZOOM INSET on subplot (b) -- SAME corner (north east) as inset (a)
  %=================================================================
  \begin{axis}[
    name=insetAxis2,
    at={(fig c1r2.north west)}, anchor=north east,
    xshift=3.25cm, yshift=-1.50cm,
    width=2.5cm, height=0.9cm,
    scale only axis=true,
    xmode=log,
    xmin=0.013, xmax=0.055, ymin=0, ymax=9.3,
    xtick={0.02,0.04},
    xticklabels={$2{\times}10^{-2}$,$4{\times}10^{-2}$},
    ytick={0,4,8},
    tick label style={font=\tiny},
    axis background/.style={fill=white, fill opacity=0.95},
    axis on top,
    axis line style={draw=gray!70},
  ]
  \addplot[rsgp10, mark repeat=1] coordinates{(0.020890,1.596)(0.050000,1.229)};
  \addplot[rsgp12, mark repeat=1] coordinates{(0.017408,3.222)(0.041667,3.123)};
  \addplot[rsgp13, mark repeat=1] coordinates{(0.016069,4.535)(0.038462,8.377)};
  \node[font=\tiny, teal!80!black, anchor=south east, inner sep=1pt]
        at (axis cs:0.020290,0.296) {$+1.60$};
  \node[font=\tiny, teal!80!black, anchor=north west, inner sep=1pt]
        at (axis cs:0.017708,5.222) {$+3.22$};
  \node[font=\tiny, teal!80!black, anchor=south west, inner sep=1pt]
        at (axis cs:0.031462,5.377) {$+8.38$};
  \end{axis}

  \draw[gray!70, thin, dashed] (zoomTL2) rectangle (zoomBR2);
  \draw[gray!50, thin] (zoomTL2) -- (insetAxis2.north west);
  \draw[gray!50, thin] (zoomBR2) -- (insetAxis2.south east);

  \end{tikzpicture}

  % \vspace{0.3em}
  \ref{sharedlegend}

  \caption{(a) Normalized classification accuracy vs.\ crossover probability $p_e$.
  (b) Normalized UEP gain $\Delta_{\mathrm{UEP}}$ vs.\ mean BER $\bar{p}$
  (fixed-ratio profile, ratio $=10$, eight representative sweep points). Marker shapes encode the bit budget ($\bullet\,m{=}10$, $\blacksquare\,m{=}12$, $\blacktriangle\,m{=}13$).}
  \label{fig:combined}
\end{figure}
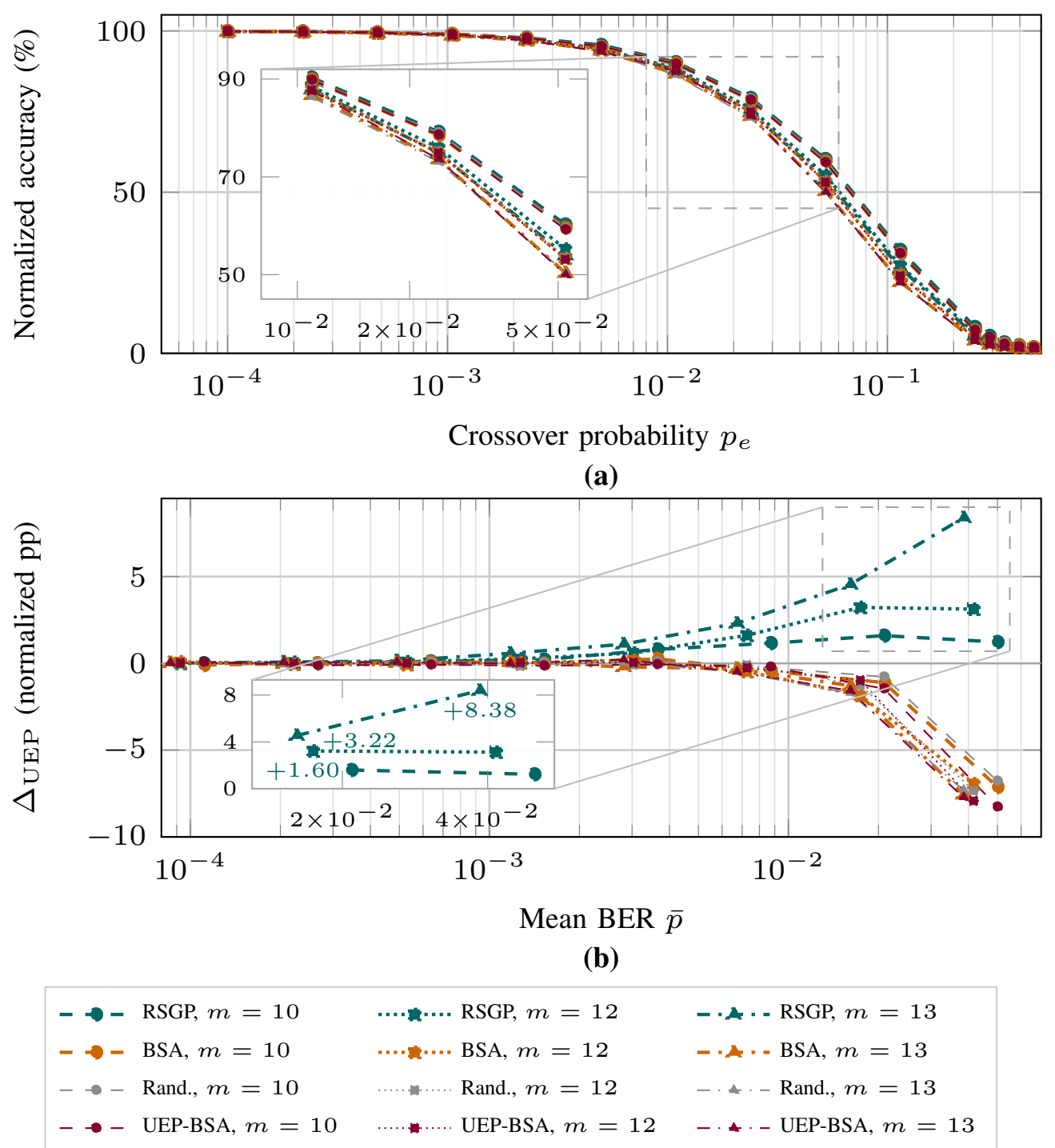

\begin{table}[!t]
  \caption{Normalized MSE at different error patterns, with the quantization-only result. Bold = lowest per $p_e$ and $m$.}
  \label{tab:mse_acc}
  \centering
  \footnotesize
  \begin{tabular}{l rrr}
    \toprule
    & $m=10$ & $m=12$ & $m=13$ \\
    \midrule
    \multicolumn{4}{l}{$p_e \approx 0.0109$} \\
    Random & 0.483 & 0.738 & 0.911 \\
    BSA    & 0.460 & 0.736 & 0.875 \\
    UEP-BSA & 0.461 & 0.746 & 0.831 \\
    RSGP & \textbf{0.409} & \textbf{0.651} & \textbf{0.747} \\
    \midrule
    \multicolumn{4}{l}{$p_e \approx 0.1143$} \\
    Random & 3.147 & 4.628 & 5.196 \\
    BSA    & 3.134 & 4.627 & 5.176 \\
    UEP-BSA & 3.136 & 4.633 & 5.197 \\
    RSGP & \textbf{2.903} & \textbf{4.230} & \textbf{4.703} \\
    \midrule
    \multicolumn{4}{l}{Quantization-only $\mathrm{MSE}_{\mathrm{quant}}$} \\
    & 2.357 & 2.135 & 2.082 \\
    \bottomrule
  \end{tabular}
\end{table}

Results in Fig.~\ref{fig:combined}(a) show that, under a uniform error profile, all mapping approaches produce close performances, even though RSGP presents a slight yet consistent normalized accuracy advantage over the entire error axis. 
This suggests that the fixed quantization scheme strongly determines the expressivity of the embedding space with respect to the label distribution, leaving only a possible yet quantization-constrained advantage in coupling IAP-based schemes to further improve classifier performance.
For $m\in\{10,12,13\}$, the resulting $K=2^m$ codebooks are likely fine-grained enough to keep transmitted embeddings close to their nearest centroids, mitigating reconstruction distortion regardless of how centroids are indexed. A clearer advantage emerges in normalized MSE (Table~\ref{tab:mse_acc}), where RSGP achieves lower reconstruction error across all $m$ and both evaluated $p_e$. Notably, RSGP's MSE gain over the other techniques matches or exceeds the gain from adding 2--3 extra quantization bits alone (cf. $\mathrm{MSE}_{\mathrm{quant}}$ for $m=10,12,13$).

These observations also motivate the study of the proposed schemes for UEP contexts, where an unequal error pattern relies on the transmitted binary representations. As shown in Fig.~\ref{fig:combined}(b), RSGP is the only method with a systematic positive gain throughout the operational range, and its maximum normalized gain increases with bit depth: approximately $+1.60$\,pp at $m=10$, $+3.22$\,pp at $m=12$, and $+8.38$\,pp at $m=13$. At high $\bar{p}$, the fixed-ratio UEP profile concentrates most errors on the least protected final bit positions, and RSGP retains a positive gain because those late bits encode fine local refinements, while Random and BSA-based mappings do not present such properties and may assign them to distinguish pairs of centroids that are still far apart. These results are also aligned with the demonstration given in Section~\ref{sec:uep-alignment}, and both experimental and analytical analyzes attest to the fact that, among the studied approaches, only the RSGP semantic-aware encoding structure can consistently reduce channel distortion against the corresponding uniform error performance over an increasing number of bits $m$. Notably, the hierarchical structure within RSGP mapping seems to be particularly exploitable under UEP and better preserves task-relevant information even for higher error probabilities (e.g. $\bar{p} \geq 10^{-2}$), where task performances tend to monotonically degrade under uniform error pattern, as indicated by Fig.~\ref{fig:combined}(a).

\subsection{Complexity Analysis}
\textbf{Random.}
To assign binary vectors for all $K$ centroids, a uniformly random permutation is generated in $\mathcal{O}(K)$ by shuffling the code matrix containing all possible $2^m$ codewords.

\textbf{BSA.}
Precomputing all pairwise squared distances costs $\mathcal{O}(K^2 d)$.
For a candidate swap of codes $\varphi(i) \leftrightarrow \varphi(j)$, the distortion change $\Delta D(i,j)$ is computed from the row-wise changes in the channel transition matrix. Evaluating a single pair means performing a XOR operation of the $K\times m$ code matrix against each row $i,j$, which gives $\mathcal{O}(Km)$ as the dominant complexity factor per pair.
A complete sweep over all $\binom{K}{2}$ candidate pairs therefore costs
\begin{equation}
  \frac{K(K-1)}{2} \cdot \mathcal{O}(Km)
  \;=\; \mathcal{O}(K^3 m)
  \;=\; \mathcal{O}(K^3 \log_2 K),
  \label{eq:bsa_sweep}
\end{equation}
since $m = \log_2 K$.
Running $T$ initialization-independent sweeps gives a total of $\mathcal{O}(K^2 d + TK^3 \log_2 K)$.

\textbf{RSGP.}
At each step of the $m$-level binary partition process, pairwise distances are recomputed on the local subproblem.
Since subproblem sizes halve at every level, the distance cost sums to $\sum_{\ell=0}^{m-1} 2^\ell \Bigl(\tfrac{K}{2^\ell}\Bigr)^{\!2} d
  = K^2 d \sum_{\ell=0}^{m-1} 2^{-\ell}
  < 2K^2 d \;=\; \mathcal{O}(K^2 d).
  \label{eq:dist_sum}$
The dominant cost at each step is the dense symmetric eigendecomposition of the local Laplacian via LAPACK's \texttt{dsyevd} (called by \texttt{numpy.linalg.eigh}), which costs $\mathcal{O}(n^3)$ for an $n\times n$ matrix~\cite{golub2013matrix}.
Summing the eigendecomposition cost over the $m$ binary partition levels: $\sum_{\ell=0}^{m-1} 2^\ell \!\cdot\! \Bigl(\tfrac{K}{2^\ell}\Bigr)^{\!3}
  = K^3 \sum_{\ell=0}^{m-1} 4^{-\ell}
  < \tfrac{4}{3} K^3 \;=\; \mathcal{O}(K^3).
  \label{eq:eigh_sum}$
The total construction cost is therefore $\mathcal{O}(K^2 d + K^3)$ with no iterative outer loop.

\section{Conclusion}
We proposed Recursive Spectral Graph Partitioning (RSGP) coding, a one-shot index assignment for quantized semantic embeddings transmitted over a channel with independent bit-flip errors.
RSGP constructs a binary mapping in $\mathcal{O}(K^2 d + K^3)$ time by recursively bisecting the centroid affinity graph along the Fiedler eigenvector, embedding a coarse-to-fine semantic hierarchy into the bit positions.
Experiments on Banking77 show that RSGP matches or exceeds BSA accuracy across different error configurations with significantly lower construction cost and is the only studied method to consistently exhibit positive UEP gain.
The latter finding highlights a new design principle: index assignments that are structurally aligned with the UEP protection gradient can exploit heterogeneous channel resources to improve task-level accuracy, independently of the distortion-minimization objective.
Future work will extend RSGP to different datasets, affinity metrics, and tasks beyond label classification.

\bibliographystyle{IEEEtran}
\bibliography{refs}

\end{document}